\documentclass{raa}

\usepackage{graphicx,times}             

\begin{document}

   \title{The progenitors of Be-stars paired with O-subdwarfs: \\the spin-up of a Be star at the stage of conservative mass exchange
}

   \volnopage{Vol.0 (20xx) No.0, 000--000}      
   \setcounter{page}{1}          

   \author{Evgeny Staritsin
   }

   \institute{K.A. Barkhatova Kourovka Astronomical Observatory, B.N. Yeltsin Ural Federal University,
             pr. Lenina 51, Ekaterinburg 620000, Russia; {\it Evgeny.Staritsin@urfu.ru}\\
   }

   \date{Received~~2009 month day; accepted~~2009~~month day}

\abstract{ The spinning-up of the accreting component in the process of conservative mass exchange is considered in binary systems - progenitors of systems consisting of a main sequence Be-star and an O-subdwarf. During the mass exchange, the meridional circulation transfers 80-85\% of the angular momentum that entered the accretor together with the accreted matter to the accretor surface. This angular momentum is removed from the accretor by the disk.  When the mass exchange finishes, the accretor has a rotation typical of classical Be-type stars.
}

   \authorrunning{E. Staritsin }            
   \titlerunning{The progenitors of Be-stars paired with O-subdwarfs }  

   \maketitle

%
%
\section{Introduction}           
\label{sect:intro}

Classical Be-stars have fast rotation (Porter \& Rivinius \cite{pr03}). The rotational velocities at the equator $V_e$ of Be-stars of early spectral subclasses (B0-B3) are in a wide range: between  $0.4\le V_e/V_c\le0.6$  at the lower limit and  $0.9\le V_e/V_c\le1.0$ at the upper limit (Cranmer \cite{c2005}), $V_c$  is the Keplerian velocity at the equator of the star. Such a rapid rotation of Be-stars could be obtained as a result of mass exchange in binary systems (Pols et al. \cite{pcwh91}; Portegies Zwart \cite{pz95}; Shao \& Li \cite{sl14}; Hastings et al. \cite{hlws21}).

To date, a number of Be-stars are found in binary systems paired with O-subdwarfs (Chojnowski et al. \cite{clrg2018}, Wang et al. \cite{wgph2023}). These binary systems have already passed the stage of mass exchange. The periods of the systems are greater than $30^d$, the masses of Be-stars are 6-12~$M_\odot$. Most Be-stars have early spectral subclasses B0-B3 (Wang et al. \cite{wgp2018}). Accreting components in the long-period Algols have increased rotation as well: $0.15\le V_e/V_c\le0.3$ (Dervisoglu et al. \cite{dti2010}). The periods of these systems are $5^d-15^d$, the masses of the accreting components are 2-7~$M_\odot$. The mass range of Be-stars in systems with O-subdwarfs partially overlaps with the range of masses of accreting components in long-period Algols. The periods of binary systems of Be-stars with subdwarfs are longer than the periods of the long-period Algols. The axial rotation of the Be-stars is also greater than the axial rotation of the accreting components in the long-period Algols.

The binary system $\phi$~Per is one of the most well-known and extensively studied systems comprising a Be-star and O-subdwarf (Gies et al. \cite{gbfk98}; Schootemeijer et al. \cite{sgmg18}). The evolution of progenitor of this system is thoroughly investigated, the system $\phi$~Per was formed as a result of conservative mass transfer
\pagebreak[4]
in the Hertzsprung-gap. Galactic Be-binaries with a helium-star companion have also been analyzed by the method of population synthesis (Shao \& Li \cite{sl21}). The orbital periods and masses of the components of these binaries can be produced due to conservative mass exchange in the Hertzsprung-gap and/or due to non-conservative mass exchange during evolution on the main sequence.
The observed characteristics of the long-period Algol systems can also be reproduced (except rotation) through simulations of conservative mass exchange
in the Hertzsprung-gap
(Van Rensbergen \& De Greve \cite{rg2020}). In the event of conservative mass accretion, the mass of the accreting component can increase by up to approximately two times.

Interacting binary stars exchange both mass and angular momentum. The axial rotation of a star receiving mass depends on the amount of angular momentum in the accreted mass, on the transfer of angular momentum in the stellar interior, and on the mechanisms of loss of angular momentum from the star. If the angular momentum is instantly redistributed in the stellar interior to solid-state rotation and there are no mechanisms for removing the angular momentum from the star, then the star can receive a critical rotation after an increase in mass by 5\%-10\%, provided that the added mass had a Keplerian rotation (Packet \cite{p1981}). A further increase in the mass and angular momentum of the star is possible due to the fact that part of the angular momentum of matter falling on the star is removed from it by an accretion disk (Paczynski \cite{p1991}; Bisnovatyi-Kogan \cite{bk93}).

In a single star, the external part with a mass equal to half the mass of the star contains more than half the angular momentum of the star (Staritsin \cite{St07}, \cite{St09}). In the case of an accreting component of a binary system and a conservative mass exchange, such an external part consists of an accreted matter. In the process of accretion, the matter brings with it the Keplerian angular momentum. This matter cannot become part of a star without decreasing the angular momentum. This decrease occurs since the meridional circulation transfers part of the angular momentum of the accreted layers to the accretor surface (Staritsin \cite{St2022}, \cite{St23a}, \cite{St2024}). This part of the angular momentum is removed from the accretor by the disk (Paczynski \cite{p1991}; Bisnovatyi-Kogan \cite{bk93}). A small part of the angular momentum of the accreted layers is transferred by the meridional circulation to the inner layers of the star.

We studied the accretor spin-up during conservative mass exchange in a binary system depending on the binary system mass and the distance between the components. The masses of binary systems and the distances between the components have values typical of the progenitors of pairs of Be-stars with O-subdwarfs and long-period binaries of the Algol type. Angular momentum transfer in accretor interrior is carried out by meridional circulation and shear turbulence.

\section{Basic equations and simplifications}
\label{sect:basic}
\subsection{The meridional circulation role in accretor spin-up}
\label{subsect:mass}

The progenitors of the binary systems with Be-stars and the long-period Algols comprise stars with radiative envelopes and convective cores. The mass exchange in the progenitor systems begins after the hydrogen exhaustion in the donor’s core. The donor fills the Roche lobe and then loses the matter in the thermal time-scale (Paczynski \cite{p1971}). The material falls into the gravitational field of the accretor swirling and forming an accretion disk around the accretor (Lubow \& Shu \cite{ls75}; Richards \& Ratliff \cite{rr1998}; Bisikalo et al. \cite{bhbk2000}; Raymer \cite{r2012}). Then the material joins the accretor, and brings with it the Keplerian moment.

The arrival of angular momentum into the upper layer of the accretor has a strong effect on the velocity field of the meridional circulation in it (Staritsin \cite{St19a}). In the upper layer of the accretor, a circulation cell of the matter is formed in the meridional plane. The circulation rate in this cell is several orders of magnitude higher than in single stars. The circulation transfers the angular momentum of the attached matter inside this cell (Staritsin \cite{St2021}, \cite{St2022}). The mass of matter in the cell increases over time for two reasons. The layers located under the cell bottom are included in the cell. New accreted layers are added from above.

After increasing the accretor mass by several percent, the rotation speed of its surface at the equator becomes equal to the Keplerian value (Staritsin \cite{St2022}, \cite{St2024}). This state of rotation of the accretor is called critical. Let $J$ and $M$ be the angular momentum and a star mass in a state of critical rotation, and $j_e^{Kep}$ be the specific Keplerian angular momentum at its equator. Then, $0<dJ/dM=j_a<j_e^{Kep}$. Accretion to a star in a state of critical rotation can continue due to the removal of excess angular momentum $\bigtriangleup j=j_e^{Kep}\!-\!j_a$ from the star by an accretion disk (Paczynski \cite{p1991}; Bisnovatyi-Kogan \cite{bk93}).
Matter rotating at Keplerian velocity joins the star. Another circulation cell is formed in the attached matter. In this cell, the circulation carries the angular momentum to the star surface (Staritsin \cite{St2022}, \cite{St2024}). The angular momentum removal from the star may occur in the accretion disk (Paczynski \cite{p1991}; Bisnovatyi-Kogan \cite{bk93}). As a result of the decrease in angular momentum, the accreted layers contract, as usually happens during the accretion process. The removal of excess angular momentum from accreted matter by meridional circulation and the removal of this angular momentum from the star by an accretion disk ensure an increase in the mass and a star angular momentum in a state of critical rotation.

\subsection{The arrival of angular momentum in the accretor}
\label{subsect:increment}

At the very beginning of the mass exchange, the angular velocity of the accretor surface is less than the Keplerian one. This results in the formation of the boundary layer between the surface of the star and the inner edge of the Keplerian accretion disk. The details of the angular momentum transfer between the boundary layer and the star, as well as between the boundary layer and the disk, do not affect the final value of the angular momentum of the accretor (Staritsin \cite{St2024}). Therefore, in current calculations, the angular velocity of the added matter is assumed to be equal to the angular velocity of the accretor surface.

At the beginning of the mass exchange, an increase in the mass of the accretor is accompanied by an increase in its size (Benson  \cite{b70}; Kippenhahn \& Meyer-Hofmeister  \cite{kmh77}). A situation may arise when a jet of matter falling from the Lagrangian point $L_1$ is directed directly or tangentially to the surface of the accretor. In this case, the disk around the accretor has a sub-Keplerian rotation (Kaitchuck \& Honeycutt  \cite{kh1982}; Kaitchuck  \cite{k1988},  \cite{k1989}; Richards \& Ratliff \cite{rr1998}; Raymer \cite{r2012}; Richards et al. \cite{rcf2014}). Lowering the disk rotation rate of below the Keplerian rotation does not affect the final value of the angular momentum of the accretor (Staritsin \cite{St2024}). Therefore, after increasing the angular velocity of the  accretor surface to the Keplerian value, the angular velocity of the added matter is assumed to be equal to the Keplerian velocity.

Angular momentum transfer in the radiative envelope of a star is carried out by meridional circulation and shear turbulence. Both angular momentum transfer processes are considered within the shellular  rotation model (Zahn \cite{Zahn92}). Angular momentum transfer is described by the principle of conservation of angular momentum (Tassoul \cite{t1978})
\begin{eqnarray}
\label{eq003}
\frac{\partial(\rho\varpi^2\Omega)}{\partial t}+
\mbox{div}(\rho\varpi^2\Omega{\bf u})
=\mbox{div}(\rho\nu_{\mbox{v}}\varpi^2\mbox{grad}\Omega).  \nonumber
\end{eqnarray}
The meridional circulation velocity $\bf u$ is determined from the law of conservation of energy in stationary form (Maeder \& Zahn \cite{mz98})
\begin{eqnarray}
\label{eq004}
\rho T{\bf u}\mbox{grad}s=\rho\varepsilon_n+
\mbox{div}(\chi\mbox{grad}T)-\mbox{div}{\bf F}_h.  \nonumber
\end{eqnarray}
In these equations,
$\Omega$ - angular velocity,
$\varpi$ - distance to the rotation axis,
$\rho$ - density,
$\nu_{\mbox{v}}$ - turbulent viscosity in the vertical direction,
$T$ - temperature,
$s$ - specific entropy,
$\varepsilon_n$ - nuclear energy release rate,
$\chi$ - thermal conductivity,
${\bf F}_h$ - turbulent enthalpy flow in the horizontal direction:
${\bf F}_h=-\nu_h\rho T\partial{s}/\partial{\bf i_\theta}$
and $\nu_h$- turbulent viscosity in the horizontal direction.
The coefficients of turbulent viscosity are  determined by Zahn (\cite{Zahn92}), Talon and Zahn (\cite{TZ97}), Maeder (\cite{m2003}), Mathis et al. (\cite{mpz04})

\begin{eqnarray}
\nu_V=\frac{2Ri}{N_T^2/(K+D_h)+N_\mu^2/D_h}
\left[\frac{d(\varpi\Omega)}{dr}\right]^2, \nonumber
\end{eqnarray}

\begin{eqnarray}
\nu_h=C\cdot r\cdot\left|U(r)\right|. \nonumber
\end{eqnarray}
Here $Ri$ - the critical Richardson number,
$N^2=N_T^2+N_\mu^2$,
$N$ - buoyancy frequency,
$K$ - thermal diffusivity,
$U(r)$  - amplitude of the vertical component of the meridional circulation velocity $U_V(r,\theta)=U(r)\cdot\mbox{P}_2(cos\,\theta)$,
$\mbox{P}_2(cos\,\theta)$ - the Legendre function of order 2.
The condition of Zahn model applicability (\cite{Zahn92}) as follows $\nu_V<\nu_h<K$.
The convective core rotates solid.

These equations are solved together with the equations of stellar structure and evolution (Staritsin \cite{St99}; \cite{St05}; \cite{St07}). Once  $U(r)$ and $\Omega(r)$ are determined, the angular momentum flux carried by circulation (advective flux) can be calculated (Zahn \cite{Zahn94})
\begin{eqnarray}
F_{ad}&=&-\frac{8\pi}{15}r^4\Omega\rho U, \nonumber
\end{eqnarray}%
angular momentum turbulent flux
\begin{eqnarray}
F_{t}&=&-\frac{8\pi}{3}\nu_{V}r^4\rho\frac{d\Omega}{dr}, \nonumber
\end{eqnarray}%
total flux
\begin{eqnarray}
F&=&F_{ad}+F_{t}. \nonumber
\end{eqnarray}%

\section{Calculation results}
\subsection{Initial parameters of the systems under study}
\begin{table}
\begin{center}
\caption[]{ Binary systems calculated parameters.}\label{Tab:publ-works}


 \begin{tabular}{ccccccccccr}
  \hline\noalign{\smallskip}
No & ${(M_d)_0}$   & ${(M_a)_0}$   & $A_0$         & $(J_a)_0$             & $\dot M$                 & ${(M_d)_f}$   & ${(M_a)_f}$   & $A_f$         & $(J_a)_f$             & counterpart   \\
   & [${M_\odot}$] & [${M_\odot}$] & [${R_\odot}$] & [$g\cdot cm^2s^{-1}$] & [${M_\odot}\,year^{-1}$] & [${M_\odot}$] & [${M_\odot}$] & [${R_\odot}$] & [$g\cdot cm^2s^{-1}$] &               \\
  \hline\noalign{\smallskip}
1  &  6            &  5            & 30            & 5.3$\times10^{50}$    & 4.5$\times10^{-5}$       & 1.5           & 9.5           & 132           & 1.83$\times10^{52}$   & $\varphi$ Per \\
2  &  3            &  2.5          & 30            & 1.4$\times10^{50}$    & 4.8$\times10^{-6}$       & 0.6           & 4.9           & 195           & 0.52$\times10^{52}$   & BD-011603     \\
3  &  3            &  2.5          & 15            & 1.4$\times10^{50}$    & 2.4$\times10^{-6}$       & 0.6           & 4.9           & 98            & 0.52$\times10^{52}$   & RX Gem        \\
  \noalign{\smallskip}\hline
\end{tabular}
\end{center}
The following is indicated in the columns:
 ${(M_d)_0}$ and ${(M_d)_f}$ - the donor initial and final mass,
 ${(M_a)_0}$ and ${(M_a)_f}$ - the accretor initial and final mass,
 $A_0$ and $A_f$ - the initial and final distance between the components,
 $(J_a)_0$ and $(J_a)_f$ - the initial and final angular momentum of the accretor,
 $\dot M$ - the average rate of mass exchange,
and  the counterpart of the calculated system.
\end{table}

We calculated the accretor spin-up during conservative mass exchange in systems with initial parameters from Table.1. The first system is similar to the progenitor of the binary system
$\phi$~Per. This system comprises a Be-star and a subdwarf O-star and is the best studied binary system of the type under consideration (Gies et al. \cite{gbfk98}; Schootemeijer et al. \cite{sgmg18}). We chose the parameters of the progenitor of this system, proposed by Vanbeveren et al. (\cite{vlr98}), in order to be able to multiply change the initial parameters of the systems under study (Table 1). The observed characteristics of the binary system $\phi$~Per are reproduced in calculations of conservative mass exchange (Schootemeijer et al. \cite{sgmg18}).
The second system has parameters close to the progenitor of BD-011603 binary system. This system also comprises a Be-star and a subdwarf O-star (Chojnowski et al. \cite{clrg2018}), but is not studied in detail. The third system (Table1) is similar to the progenitor of the long-period Algol-type system RX Gem (Van Rensbergen \& De Greve \cite{rg2020}). The observed parameters of the long-period Algols can also be reproduced (except for rotation) in calculations of conservative mass exchange.

The accretion rate is assumed to be equal to the average value. The accretor mass increase is equal to the difference between the initial mass of the donor and the mass of the remnant. The mass of remnant can be determined to the formula Massevitch and Tutukov (\cite{mt1988}) as follows
\begin{eqnarray}
(M_d)_f=\zeta(M_d)_0^{1.4}. \nonumber
\end{eqnarray}%
The coefficient $\zeta$ depends on the input physics,  the initial chemical composition, and the overshooting parameter. Here this coefficient is determined using calculations by Vanbeveren et al. (1998) for the binary system $\phi$~Per. The mass exchange duration is a tripled thermal time-scale $t_{KH}$ (Paczynski \cite{p1971}), where
\begin{eqnarray}
t_{KH}=3.12\times10^7\frac{(M_d)_0^2}{(R_d)_0(L_d)_0}, \nonumber
\end{eqnarray}%
$(R_d)_0$ and $(L_d)_0$ are the size and luminosity of the donor before mass exchange. The average values of the accretion rate are shown in Table 1.

   \begin{figure}
   \centering
   \includegraphics[width=\textwidth, angle=0]{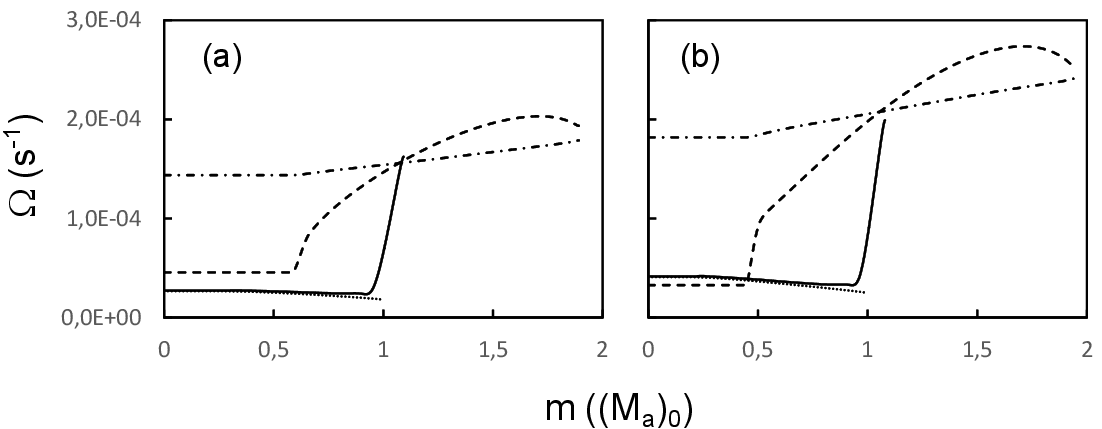}
   \caption{
   The angular velocity distribution in the accretor interrior prior to the mass exchange (dotted line), when the angular velocity of the surface increases to the Keplerian value (solid line), after the mass exchange (dashed line), after the thermal equilibrium restoration (dot-dashed line). The distributions are shown for the accretor from the first (a) and second (b) systems in Table 1.
   }
   \label{Fig1}
   \end{figure}

The initial distance between the components is the smallest in the third system shown in Table.1.
The synchronization time of axial rotation with orbital motion (Zahn \cite{Zahn75}; Hurley et al. \cite{htp02}) for a less massive star in this system is less than the evolution time of a more massive star on the main sequence.
Therefore, the less massive star acquires
synchronous rotation before the mass exchange begins. The angular velocity is $\sim$10\% of the Keplerian value. In the first and second systems shown in Table.1 component with a lower mass may have a rotation other than synchronous.  For the uniformity of the initial data, the angular velocity of the components surface in these systems is assumed to be 10\% of the Keplerian velocity. In all cases, the accretor angular momentum prior to the start of mass exchange $J_0$ has a small value (Table 1). The meridional circulation rate is $\sim10^{-6}$cm/s in a star with a mass of 5~$M_\odot$ and $\sim10^{-7}$cm/s in a star with a mass of 2.5~$M_\odot$. Angular momentum transfer in the interiors of stars is not effective.

The hydrogen content in the convective cores of a component with masses of 5 $M_\odot$ and 2.5 $M_\odot$ before mass exchange is ~0.4. During the accretion process, the cores mass and the hydrogen content increase. To simplify calculations, we assumed that the distribution of hydrogen in the convective core and the layer above is the same as that of a single star with the same mass and amount of helium as that of an accretor.

   \begin{figure}
   \centering
   \includegraphics[width=\textwidth, angle=0]{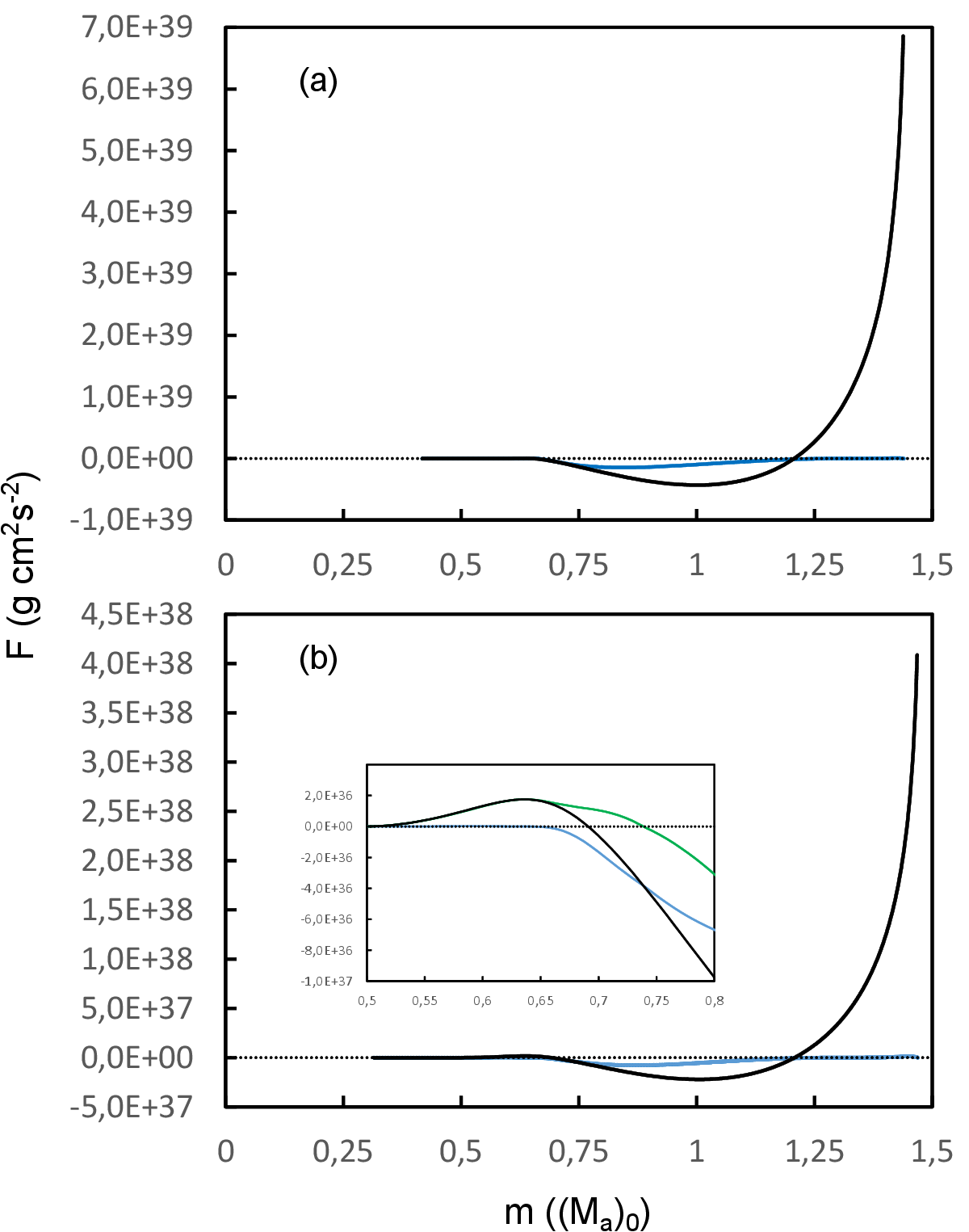}
   \caption{
   The angular momentum flux $F$ inside the accretor after attaching to it the first half of the mass lost by the donor (black solid line), as well as the turbulent flux of angular momentum $F_t$ (blue solid line). The distributions are shown for the accretor from the first (a) and second (b) systems in Table.1. The advective angular momentum flux $F_{ad}$ (green solid line) is shown in the box.
   }
   \label{Fig2}
   \end{figure}

\subsection{Accretor spin-up in the systems with $A_0=30\,R_\odot$}

   \begin{figure}
   \centering
   \includegraphics[width=\textwidth, angle=0]{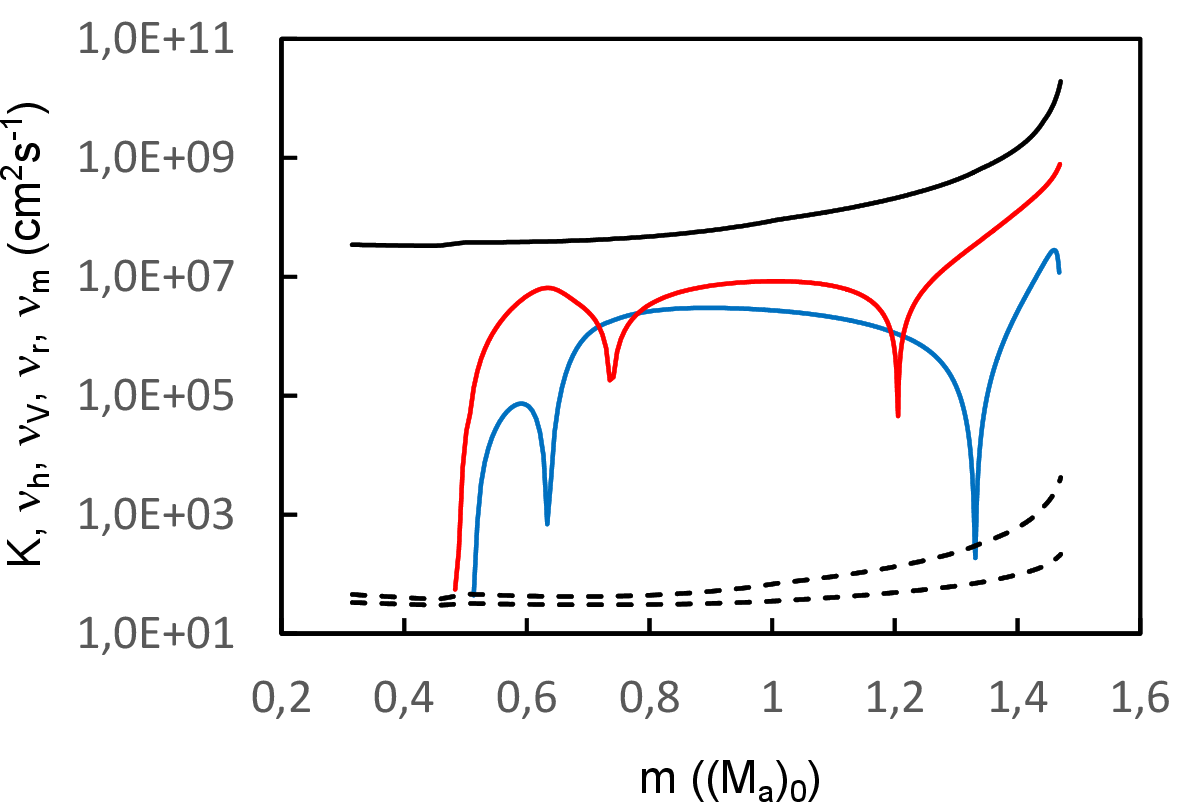}
   \caption{
   The thermal diffusivity $K$ (black solid line), turbulent viscosity in the horizontal $\nu_h$ (red solid line) and vertical $\nu_V$ (blue solid line) directions, the viscosity of radiation $\nu_r$ (upper dashed line) and matters $\nu_m$ (lower dashed line) inside the accretor from the second system Table.1 after the first half of the mass lost by the donor is attached to it.
   }
   \label{Fig3}
   \end{figure}

In the cases of both systems (Table 1), a new meridional circulation cell is formed at the beginning of mass exchange in the accreted layers and the layers located below. The maximum value of the meridional circulation velocity in this cell is $8\times10^{-2}$~cm/s in the case of an accretor from the first system (see Table 1) and $8\times10^{-3}$~cm/s in the case of an accretor from the second system. The circulation transfers the angular momentum of the accreted matter into the cell. A cascade of cells with the opposite direction of angular momentum transfer is formed below. The maximum circulation velocity in these cells is significantly lower. The maximum velocity is lower the lower the cell is located. In the cascade cells, there is a slight redistribution of the initial angular momentum between the layers in which this cascade is formed. As the mass of the accretor increases, the angular velocity of its surface increases. After an increase in the accretor mass by $\sim$10\%, the surface angular velocity is compared with the Keplerian value (Fig.1).

Once the angular velocity of the accretor surface is equal the Keplerian rotation, another circulation cell is formed in the newly accreted layers. In this cell, the circulation transfers part of the angular momentum of the accreted matter to the accretor surface. This part of the angular momentum is removed from the star by a disk (Paczynski \cite{p1991}; Bisnovatyi-Kogan \cite{bk93}). Thus, the angular momentum of the accreted layers decreases and these layers contract. The angular velocity of the accretor surface remains equal to the Keplerian value all the time.

The main role of the meridional circulation is to transfer part of the angular momentum of the accreted layers to the accretor surface at the stage of mass exchange when the angular velocity of the accretor surface is equal to the Keplerian value. The outward flux of angular momentum in the outer circulation cell is due exclusively to the large-scale flow of matter (Fig. 2). The maximum velocity of meridional circulation in this cell is 0.9 cm/s (for the accretor from the first system, Table.1) and 1.2 cm/s (for the accretor from the second system). Note that the angular velocity in the lower part of the cell increases outward, and decreases in the upper part (Fig.1).

The transfer of angular momentum inside the accretor continues in the circulation cell, which was formed at the beginning of mass exchange (Fig. 2). The maximum velocity of meridional circulation in this cell is $2\times10^{-2}$~cm/s (for the accretor from the first system, Table.1) and $2\times10^{-3}$~cm/s (for the accretor from the second system). The contribution of turbulence to the angular momentum transfer varies from insignificant in the upper part of the cell to the main one at the lower boundary of the cell. The coefficient of turbulent viscosity in the horizontal direction $\nu_h$ is determined at $C=1/20$.The coefficient has two minima at the boundaries of the considered circulation cell (Fig.3). The coefficient of turbulent viscosity in the vertical direction $\nu_V$ has one minimum in the outer circulation cell, where the angular velocity has a maximum, and another located below the considered cell.

At the lower boundary of the cell under consideration, the advective flux of angular momentum $F_{ad}$ turns to zero. However, the turbulent flux $F_t$ is different from zero and is directed inside the accretor (insert in Fig.2b). Due to this, the layers located below the lower boundary of the cell spin-up as well. These layers are then attached to the circulation cell. Turbulence contributes to the lowering of the lower boundary of this cell into the accretor. The second minimum in the distribution of the turbulent viscosity coefficient $\nu_V$ (Fig.3) is located at the intersection of the swirled and non-swirled layers of the accretor. Below, in one of the cascade cells, there is a slight redistribution of the initial angular momentum between the layers trapped in this cell. The turbulent viscosity in these layers decreases to the values of the radiation viscosity $\nu_r$ and the viscosity of the matter $\nu_m$ (Fig.3).

The fraction per mass of the matter swirled due to the transfer of angular momentum from the accreted matter increases over time in the same way in accretors with different initial masses. However, the fraction per mass of the convective core is greater for the accretor with a larger mass. The lower boundary of the cell, in which circulation and turbulence transfer angular momentum into the star, descends to the convective core in an accretor with an initial mass of 5 $M_\odot$ when the mass exchange is not ended. In this case, angular momentum begins to flow into the convective core during mass exchange. In an accretor with an initial mass of 2.5 $M_\odot$, the lower boundary of the cell descends to the convective core at the very end of the mass exchange (Fig.1). Angular momentum begins to flow into the convective core after the end of the mass exchange.

After the end of mass exchange, 18\% of the angular momentum brought with the accreted matter in the case of the first system and 16\% in the case of the second remains in the accretor.
   \begin{figure}
   \centering
   \includegraphics[width=\textwidth, angle=0]{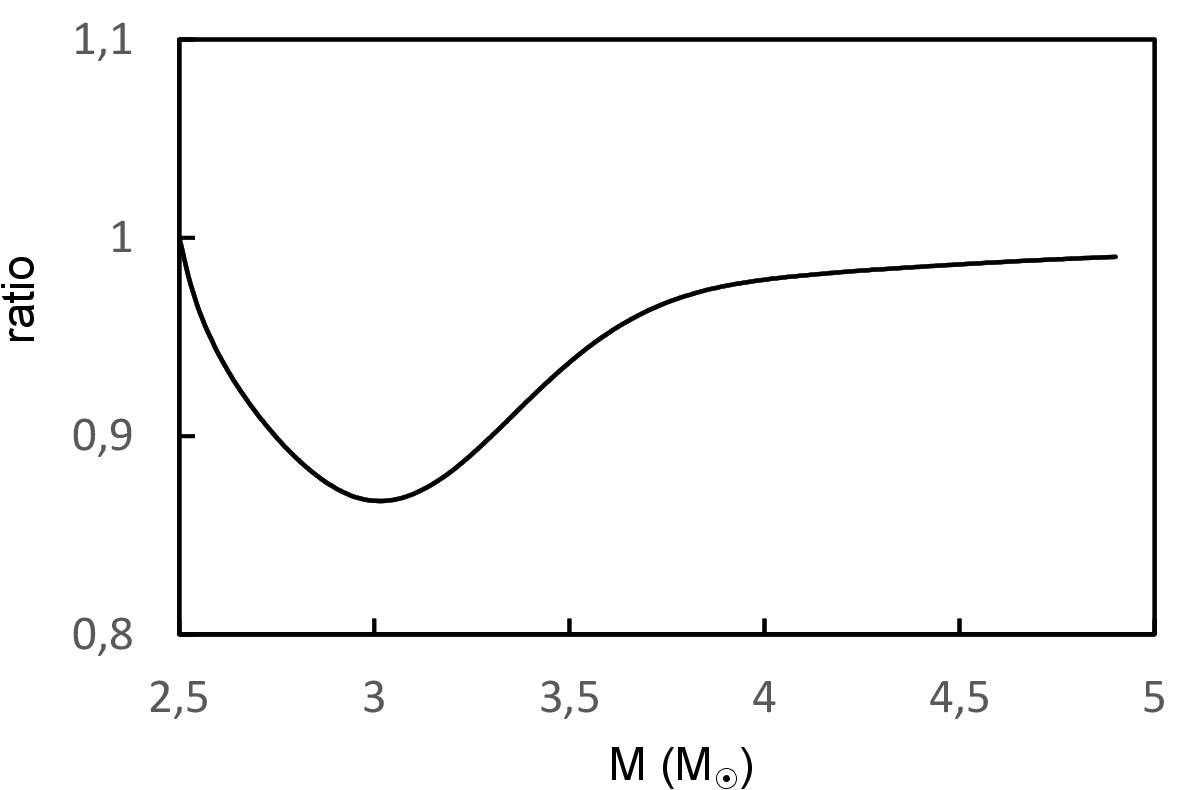}
   \caption{
   The ratio of the accretor size from the third system of Table 1 to the size of the accretor from the second system during mass exchange, depending on the accretor mass.
   }
   \label{Fig4}
   \end{figure}

   \begin{figure}
   \centering
   \includegraphics[width=\textwidth, angle=0]{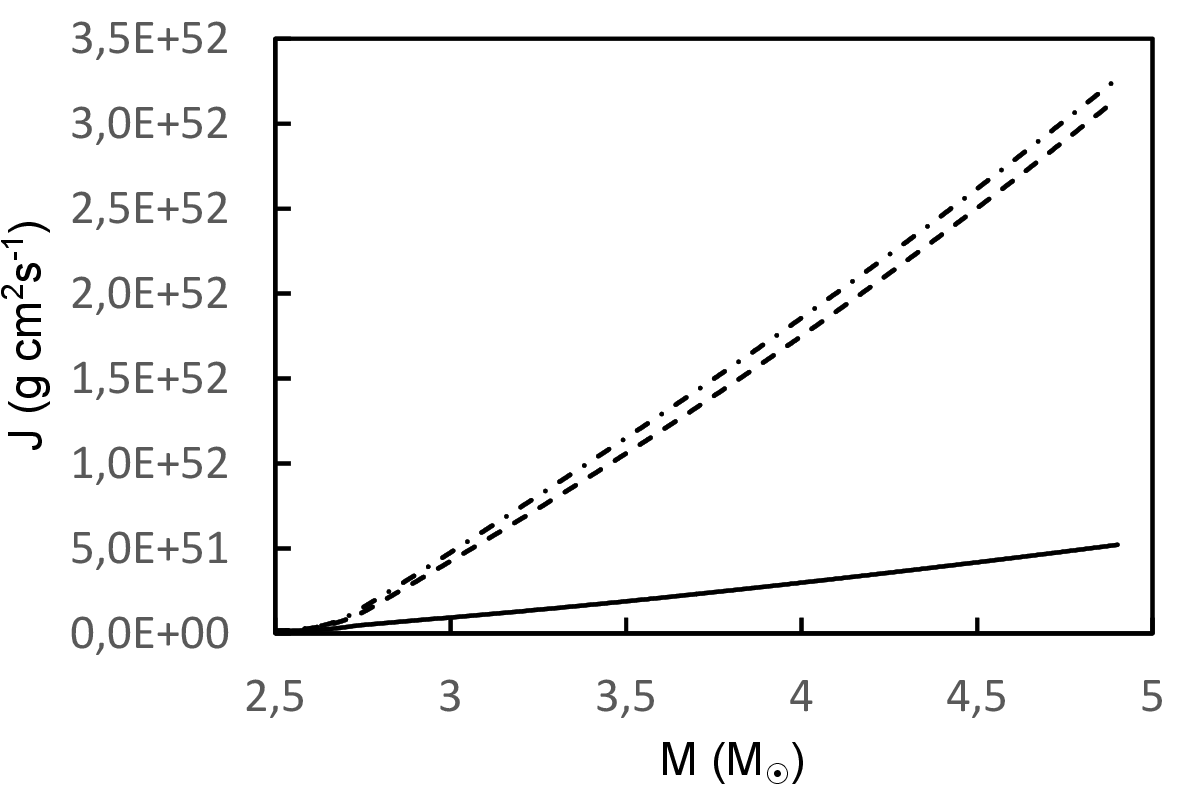}
   \caption{
   The angular momentum brought by the matter into the accretor during mass exchange in the second (dot-dashed line) and third (dashed line) systems from Table.1, as well as the accretor angular momentum  (solid line) in these systems, depending on the accretor mass.
   }
   \label{Fig5}
   \end{figure}

\subsection{Accretor spin-up in the systems with $(M_a)_0=2.5\,M_\odot$}

The thermal time-scale of a star depends on its size. The duration of mass exchange is longer, the rate of mass exchange and the average value of the rate are smaller in a system with a smaller distance between the components (Table 1). The  accretor size increases during accretion less in the case when the rate of accretion is less. The ratio of the accretor size in the cases $A_0=15\, R_\odot$ and $A_0=30\, R_\odot$ is shown in Fig.4. The matter is added to the accretor with a Keplerian rotational velocity. Therefore, the angular momentum brought into the accretor by the matter in the case of $A_0=15\, R_\odot$ is less than in the case of $A_0=30\, R_\odot$. Nevertheless, the accretor angular momentum after the end of mass exchange is the same in the cases $A_0=15\, R_\odot$ and $A_0=30\, R_\odot$ (Fig.5). The angular momentum transferred by circulation to the surface of the accretor and then lost by the accretor depends on the initial distance between the components.

The conclusion about the independence of the final angular momentum of the accretor from the angular velocity of the accreted matter was obtained previously in calculations where the angular velocity of the accreted matter varied (Staritsin \cite{St23b}, \cite{St2024}).

\subsection{Evolution of angular velocity profile after mass exchange}

After the end of the mass exchange, the accretor restores thermal equilibrium. A decrease in the size of the accretor while retaining angular momentum would result in an excess of the surface angular velocity of the Kepler value. The restoration of the thermal equilibrium of the accretor occurs with a slight loss of angular momentum. The surface angular velocity retains the Keplerian value. The need to reduce angular momentum may be the reason for the formation of a deccretion disk immediately after the end of mass exchange. Such a binary system can be observed as a system with a disk and without a matter flow between components.

The main process at the stage of restoring the accretor thermal equilibrium is the  angular momentum transfer into its inner layers (Fig.6). As a result, the angular velocity of the convective core increases, and the angular velocity of the outer part of the radiative envelope decreases (Fig.1). Such a change in angular velocity contributes to a decrease in the velocity of meridional circulation. By the end of the thermal equilibrium restoration, the maximum value of the circulation velocity is $9\times10^{-3}$~cm/s in an accretor with a final mass of 9.5 $M_\odot$ and $10^{-3}$~cm/s in an accretor with a final mass of 4.9 $M_\odot$.

Over a longer period, evolutionary changes in the structure of the star affect the change in angular velocity. Slight compression of the core and the envelope expansion contribute to an increase in the core angular velocity and a decrease in the envelope angular velocity. With the continued transfer of angular momentum into the star, the core begins to rotate faster than the envelope. Soon after, the direction of angular momentum transfer changes, and circulation and turbulence begin to transfer angular momentum outward, as happens in rotating stars of the main sequence.

   \begin{figure}
   \centering
   \includegraphics[width=\textwidth, angle=0]{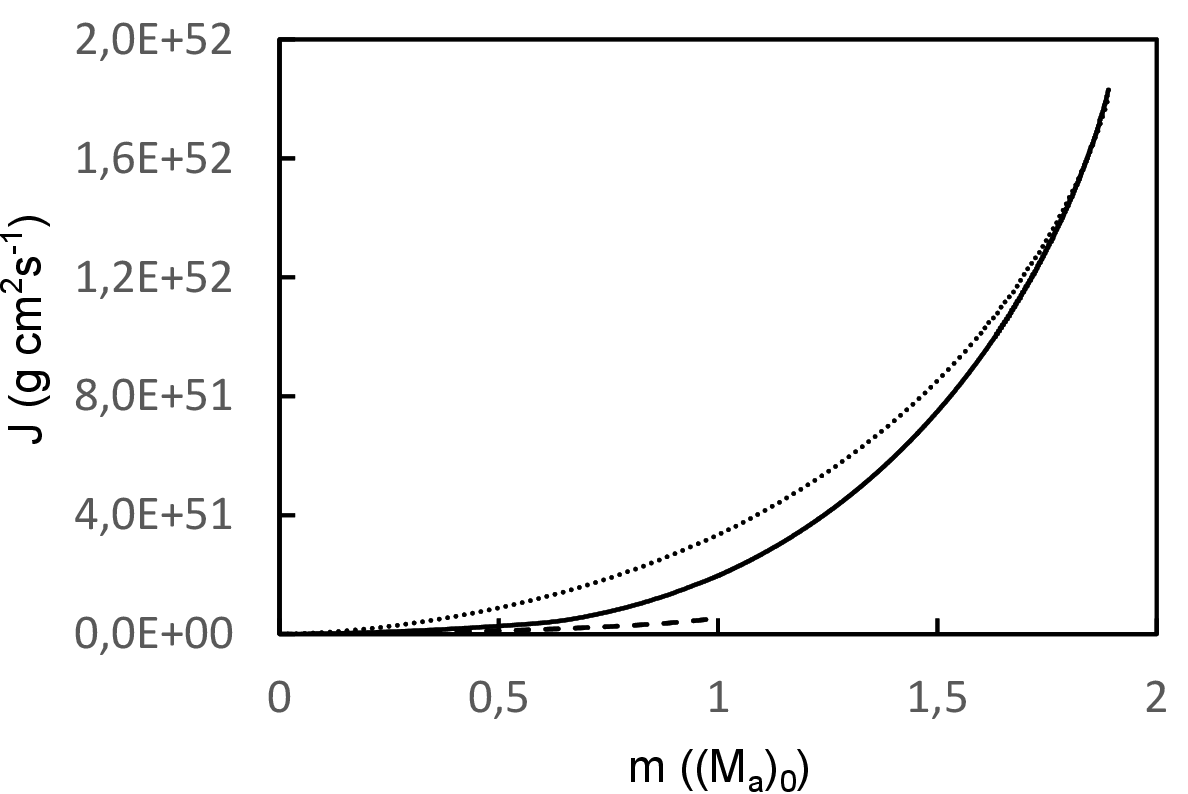}
   \caption{
   The angular momentum distribution in the accretor interior from the first system of Table 1 prior to the mass exchange (dashed line), at the end of the mass exchange (solid line) and after the thermal equilibrium restoration (dotted line).
   }
   \label{Fig6}
   \end{figure}

\section{Discussion and conclusions}

The mass of the accretor convective core increases during mass exchange. In all the considered cases, the hydrogen content in the convective core increases to $\sim$0.6 by the end of mass exchange. The accretor receives the characteristics of a slightly evolved main sequence star. The moment of inertia of the accretor is 14\% greater than that of a zero-age star with the same mass. In principle, the accretor angular momentum may be greater than the maximum angular momentum of the zero-age star model. But due to the removal of part of the angular momentum from the accretor by circulation and the disk during mass exchange, the angular momentum of the accretor turns out to be less than the maximum angular momentum of the zero-age star model. On the other hand, the accretor receives a faster rotation than is observed in young stars of the early spectral subclass (B0-B3). The evolution of stars with such angular momentum (Table.1) is accompanied by an increase in the $V_e/V_c$ ratio over time (Meynet \& Maeder \cite{mm05}; Staritsin \cite{St07}; Ekstrom et al. \cite{emmb08}; Granada et al. \cite{geg13}). An accretor may have the characteristics of a Be-star during subsequent evolution in a binary system if the distance between the components is large enough for the tidal interaction to be negligible. Thus, the conservative mass exchange in a binary system may be the reason for the rapid rotation of Be-stars over the entire mass range of stars of this type observed in pairs with O-subdwarfs.

The method of population synthesis makes it possible to trace the evolution of a large number of binary systems by means of a simplified description of the stage of mass exchange. In particular, the rigid rotation of the accreting star is considered (Mink et al. \cite{mli13}). This is due to the slight difference in the rotation of single main sequence stars from rigid rotation. Detailed calculations of angular momentum transfer in the interior of the accreting star showed differential rotation in its interior. After the end of the mass exchange and the restoration of thermal equilibrium, the rotation of the accretor is not much different from the rigid. In the case of conservative mass exchange, the mass of the accreted matter is about half the mass of the accretor. The transfer of angular momentum from the accreted matter to the interior of the star during the restoration of thermal equilibrium does little to reduce the surface angular velocity. Therefore, the assumption of rigid rotation of the accreting star in the method of population synthesis is justified in the case of conservative mass exchange.

The accretor spin-up in the case of conservative mass exchange in the Hertzsprung gap does not depend on the initial distance between the binary system components. The removal of part of the angular momentum from the accretor by the circulation and the disk during mass exchange is not able to explain the observed rotation of the accreting components of long-period Algols. The shear flow that forms inside the accretor during mass exchange can be a source of a magnetic field. Under the suitable conditions, the interaction of the magnetic field and the disk can decelerate the accretor rotation. The effect of this mechanism does not depend on the initial distance between the binary system components. Magnetic field deceleration is also unable to explain the observed rotation of accreting components of the long-period Algols (Van Rensbergen \& De Greve \cite{rg2020}). The only deceleration mechanism, the effect of which depends on the distance between the binary system components is the tidal interaction. Perhaps studying the details of this mechanism will help solve the rotation issue of accreting components in the long-period Algols.

\begin{acknowledgements}
This work was supported by the Ministry of Science and Education, FEUZ-2020-0038
\end{acknowledgements}

%

\label{lastpage}


\begin{thebibliography}{99}


  \bibitem[1970]{b70} Benson, R. S. 1970, BAAS, 2, 295

  \bibitem[2000]{bhbk2000} Bisikalo, D.V., Harmanec, P., Boyarchuk, A.A., Kuznetsov, O.A., \& Hadrava, P. 2000, \aap, 353, 1009

  \bibitem[1993]{bk93} Bisnovatyi-Kogan, G.S. 1993, \aap, 274, 796

  \bibitem[2018]{clrg2018} Chojnowski, S.D., Labadie-Bartz, J., Rivinius, T., et al. 2018, \apj, 865, 76

  \bibitem[2005]{c2005} Cranmer, S.R. 2005, \apj, 634, 585

  \bibitem[2010]{dti2010} Dervisoglu, A., Tout, C.A., \& Ibanoglu, C. 2010, \mnras, 406, 1071

  \bibitem[2008]{emmb08} Ekstrom, S., Meynet, G., Maeder, A., \& Barblan, F. 2008, \aap, 478, 467

  \bibitem[1998]{gbfk98} Gies, D.R., Bagnuolo, W.G., Ferrara, E.C., et al. 1998, \apj, 493, 440

  \bibitem[2013]{geg13} Granada, A., Ekstrom, S., Georgy, C. et al. 2013, \aap, 553, A25

  \bibitem[2021]{hlws21} Hastings, B., Langer, N., Wang, C., Schootemeijer, A., \& Milone, A.P. 2021, \aap, 653, A144

  \bibitem[2002]{htp02} Hurley, J.R., Tout, C.A., \& Pols, O.R. 2002, \mnras, 329, 897

  \bibitem[1988]{k1988} Kaitchuck, R.H. 1988, \pasp, 100, 594

  \bibitem[1989]{k1989} Kaitchuck, R.H. 1989, \ssr, 50, 51

  \bibitem[1982]{kh1982} Kaitchuck, R.H., \& Honeycutt, R.K. 1982,  \apj, 258, 224

  \bibitem[1977]{kmh77} Kippenhahn, R., \& Meyer-Hofmeister, E. 1977, \aap, 54, 539

  \bibitem[1975]{ls75} Lubow, S.H., \& Shu, F.H. 1975, \apj, 198, 383

  \bibitem[2003]{m2003} Maeder, A. 2003, \aap, 399, 263

  \bibitem[1998]{mz98} Maeder, A., \& Zahn, J.-P. 1998, \aap, 334, 1000

  \bibitem[1988]{mt1988} Massevitch, A.G., \& Tutukov, A.V. 1988 Evolution of Stars: theory and observation (Nauka, Moscow)

  \bibitem[2004]{mpz04} Mathis, S., Palacios, A., \& Zahn, J.-P. 2004, \aap, 425, 243

  \bibitem[2005]{mm05} Meynet, G., \& Maeder, A. 2005, ASP Conf. Ser., 337, 15 p.

  \bibitem[2013]{mli13} Mink, S.E., Langer, N., Izzard, R.G., Sana, H., \& de Koter, A. 2013, \apj, 764, 166

  \bibitem[1981]{p1981} Packet, W. 1981, \aap, 102, 17

  \bibitem[1971]{p1971} Paczynski, B. 1971, \araa, 9, 183

  \bibitem[1991]{p1991} Paczynski, B. 1991, \apj, 370, 597

  \bibitem[1991]{pcwh91} Pols, O.R., Cote, J., Waters, L.B.F.M., \& Heise, J. 1991, \aap, 241, 419

  \bibitem[1995]{pz95} Portegies Zwart, S.F. 1995, \aap, 296, 691

  \bibitem[2003]{pr03} Porter, J.M., \& Rivinius, T. 2003, \pasp, 115, 1153

  \bibitem[2012]{r2012} Raymer, E. 2012, \mnras, 427, 1702

  \bibitem[2014]{rcf2014} Richards, M.T., Cocking, A.S., Fisher, J.G., \& Conover, M.J. 2014, \apj, 795, 160

  \bibitem[1998]{rr1998} Richards, M.T., \& Ratliff, M.A. 1998, \apj, 493, 326

  \bibitem[2018]{sgmg18} Schootemeijer, A., Gotberg, Y., de Mink, S.E., Gies, D., \& Zapartas, E. 2018, \aap, 615, A30

  \bibitem[2014]{sl14} Shao, Y., \& Li, X.-D. 2014, \apj, 796, 37

  \bibitem[2021]{sl21} Shao, Y., \& Li, X.-D. 2021, \apj, 908, 67

  \bibitem[2021]{St2021} Staritsin, E. 2021, \aap, 646, A90

  \bibitem[2022]{St2022} Staritsin, E. 2022, RAA, 22, 105015

  \bibitem[2024]{St2024} Staritsin, E. 2024, RAA, 24, 015001

  \bibitem[1999]{St99} Staritsin, E.I. 1999, Astronomy Reports, 43, 592

  \bibitem[2005]{St05} Staritsin, E.I. 2005, Astronomy Reports, 49, 634

  \bibitem[2007]{St07} Staritsin, E.I. 2007, Astronomy Letters, 33, 93

  \bibitem[2009]{St09} Staritsin, E.I. 2009, Astronomy Letters, 35, 413

  \bibitem[2019]{St19a} Staritsin, E.I. 2019, \apss, 364, 110

  \bibitem[2023a]{St23a} Staritsin, E.I. 2023a, Astronomy Reports, 67, 959

  \bibitem[2023b]{St23b} Staritsin, E.I. 2023b, INASAN Science Reports, 8, 54

  \bibitem[1997]{TZ97} Talon, S., \& Zahn, J.-P. 1997, \aap, 317, 749

  \bibitem[1978]{t1978} Tassoul, J.-L. 1978 Theory of Rotating Stars (Princeton Univ. Press)

  \bibitem[2020]{rg2020} Van Rensbergen, W., \& De Greve, J.P. 2020, \aap, 642, 183A

  \bibitem[1998]{vlr98} Vanbeveren, D., De Loore, C., \& Van Rensbergen, W. 1998, \aapr, 9, 63

  \bibitem[2018]{wgp2018} Wang, L., Gies, D.R., \& Peters, G.J. 2018, \apj, 853, 156

  \bibitem[2023]{wgph2023} Wang, L., Gies, D.R., Peters, G.J., \& Han, Z. 2023 \aj, 165, 203

  \bibitem[1975]{Zahn75} Zahn, J.-P. 1975, \aap, 41, 329

  \bibitem[1992]{Zahn92} Zahn, J.-P. 1992, \aap, 265, 115

  \bibitem[1994]{Zahn94} Zahn, J.-P. 1994, \ssr, 66, 285






































\end{thebibliography}
\end{document}